\documentclass{article}

\PassOptionsToPackage{numbers, compress}{natbib}

\usepackage[final]{neurips_2019_ml4ps}

\usepackage[utf8]{inputenc} 
\usepackage[T1]{fontenc}    
\PassOptionsToPackage{hyphens}{url}
\usepackage{hyperref}       
\usepackage{booktabs}       
\usepackage{amsfonts}       
\usepackage{nicefrac}       
\usepackage{microtype}      
\usepackage{amsmath}
\usepackage{graphicx} 
\usepackage{multirow}
\usepackage{array}
\usepackage{caption}
\usepackage{subcaption}
\usepackage{amssymb}

\title{Learning to Reconstruct Crack Profiles for Eddy Current Nondestructive Testing}

\author{%
Shaohua Li \hspace{5pt}
Ayesha Anees \hspace{5pt} Yu Zhong \hspace{5pt}
Zaifeng Yang \hspace{5pt} 
Yong Liu \hspace{5pt} \vspace{3pt} \\
\textbf{Rick Siow Mong Goh} \hspace{5pt} \textbf{En-Xiao Liu} \vspace{8pt} \\ 
Institute of High Performance Computing, A*STAR, Singapore\\
\{li\_shaohua, ayesha\_anees, zhongyu, yang\_zaifeng, liuyong, gohsm, liuex\}@ihpc.a-star.edu.sg \\
}

\begin{document}

\maketitle

\begin{abstract}
Eddy current testing (ECT) is one of the most popular Nondestructive Testing (NDT) techniques, especially for conductive materials. Reconstructing the crack profile from measured EC signals is one of the main goals of ECT. This task is highly challenging, as the EC signals are nonlinear responses resulted from the presence of cracks, and reconstructing the crack profile requires establishing the forward model of the nonlinear electromagnetic dynamics and solving its inverse problem, which is an ill-posed numerical optimization problem. Instead of solving the inverse problem numerically, we propose to directly learn the inverse mapping from EC signals to crack profiles with a deep encoder-decoder convolutional neural network named \emph{EddyNet}. EddyNet is trained on a set of randomly generated crack profiles and the corresponding simulated EC responses generated from a realistic forward model. On the held-out test data, EddyNet achieved a mean absolute error of 0.198 between predicted profiles and ground truth ones. Qualitatively, the geometries of predicted profiles are visually similar to the ground truth profiles. Our method greatly reduces the usual reliance on domain experts, and the reconstruction is extremely fast both on GPUs and on CPUs. The source code of EddyNet is released on \url{https://github.com/askerlee/EddyNet}.
\end{abstract}

\section{Introduction}
Nondestructive testing (NDT) is a group of techniques to inspect materials, components or systems for discontinuities without causing damage of the inspected specimen. One of the most popular NDT methods is Eddy current testing (ECT), which is extensively used for surface inspections and tubing inspections of conductive materials in the aerospace and petrochemical industries. Conventional ECT is limited in that it only identifies the existence of defects within the specimen, but does not tell further information about the found defects, such as their geometries (sizes and shapes). As such information could be vital for structural health monitoring and life time prediction, imaging methods using Eddy current (EC) signals have been studied in recent years \cite{Hughes_2015, Javier_ASTNE_2011}. Most existing methods adopt optimization techniques to find the profiles of defects \cite{Bowler_AIP_2002}. However, it is a typical nonlinear and ill-posed inverse problem. Finding its solution requires sophisticated optimization techniques and takes many minutes even for a small scale problem\footnote{The forward problem, i.e., computing the impedance given a crack profile, can be solved within seconds using techniques such as the Boundary Element Method \cite{Miorelli_TM_2013}.}.

In this paper, contrary to the conventional paradigm of solving the inverse problem numerically, we propose to directly learn the inverse mapping from EC signals to crack profiles with a deep encoder-decoder convolutional neural network, named \emph{EddyNet}.

\begin{figure}[th]
\centering
\begin{minipage}{0.4\textwidth}
\includegraphics[width=\textwidth,scale=0.7]{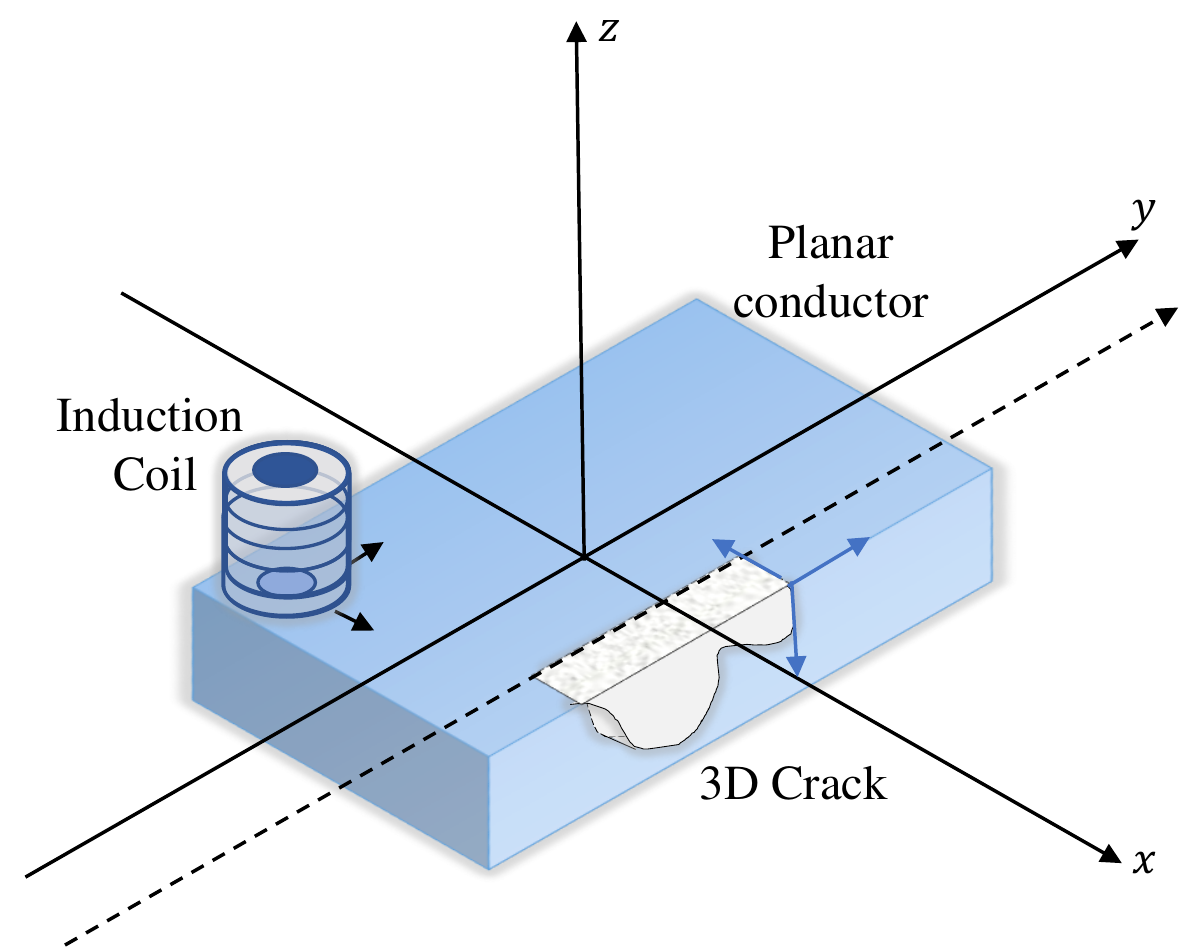}
\captionof{figure}{Eddy current testing on a crack varying along the $(y,z)$ plane and uniform along the $x$ axis.}
\label{setup}
\end{minipage}\hspace{1.5em}
\begin{minipage}{0.55\textwidth}
\centering
\newcolumntype{C}{ >{\centering\arraybackslash} m{1.1cm} }
\begin{tabular}{CCCCC}
\scriptsize{Crack Profile} & \scriptsize{Frequency 1} & \scriptsize{Frequency 2} & \scriptsize{Frequency 3} & \tabularnewline
\tabularnewline
\multirow{2}{*}{\includegraphics[viewport=0bp -8bp 12bp 32bp,scale=2]{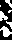}} & \includegraphics{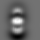} & \includegraphics{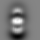} & \includegraphics{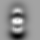} & \scriptsize{Real Parts}\tabularnewline
 & \includegraphics{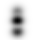} & \includegraphics{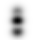} & \includegraphics{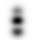} & \scriptsize{Imaginary Parts}\tabularnewline
\end{tabular} 
\captionof{figure}{A randomly generated binary crack profile and the corresponding simulated EC responses in the frequency domain.}
\label{data-sample}
\end{minipage}
\end{figure}

An EddyNet consists of an \emph{encoder} and a \emph{decoder}. The encoder takes a few channels of 2D EC responses (at different frequencies) as input, transforms them through multiple convolutional layers, yielding a continuous vector of \emph{latent code} to represent the input. The decoder transforms the latent code through multiple deconvolutional layers, and outputs a reconstructed crack profile.

In order to train an EddyNet, a realistic forward model is established to generate a set of random crack profiles and their corresponding simulated EC responses. The reconstruction errors on these (profile, responses) pairs are used as supervision signal to train EddyNet parameters.

Similar methods of approaching inverse problems using CNNs have been proposed in \cite{inv-unet, linear} (see \cite{review} for a review). However these works assume that the forward problems are linear. In contrast, EddyNet is exempt from such limitations and thus has much broader potential applications.

\section{Physical Background and Data Simulation} \label{background}

This section presents the background knowledge of the \emph{forward model} for ECT with a coil enhanced with a Ferrite core.

In ECT, we typically investigate specimens with layered structures. The coil is placed above such specimens and moved along the $x,y$ axes to obtain EC responses. The system setup is illustrated in Fig. \ref{setup}, where for simplicity, we assume that the 3D crack has a uniform width along the $x$ axis, and we only need to estimate the crack profile in the $(y,z)$ plane.

The impedance of the coil can be calculated via (\cite{Sabbagh_TM_1987, Bowler_TM_1989}): 
\vspace{-2pt}
\begin{equation}
\label{eq_Z}
    Z = -\frac{1}{I} \int_{\rm coil} \bar{E}(\bar{r}) \cdot \bar{J}(\bar{r}) d\bar{r},
\end{equation}\vspace{-2pt}%
where $\bar{E}$ is the total fields and $\bar{J}$ is the current density of the hard-source current on the coil. It can be seen from Eq.~\eqref{eq_Z} that, if defects exist, the total electric fields change due to the influences of the defects, resulting in a change of the impedance. The impedance variation may be calculated through 
\vspace{-2pt}
\begin{equation}
    \Delta Z = -\frac{1}{I} \int_{\rm coil} \Delta \bar{E}(\bar{r}) \cdot \bar{J}(\bar{r}) d\bar{r}.
\end{equation}\vspace{-2pt}%
The equation above can be rewritten, by the reciprocal theorem, as
\vspace{-2pt}
\begin{equation}
\label{Z_variation}
    \Delta Z = -\frac{1}{I} \int_{\substack{\rm crack\\ \rm profile}} \bar{E}^{\rm inc}(\bar{r}) \cdot \bar{J}^{\rm ind}(\bar{r}) d\bar{r},
\end{equation}\vspace{-2pt}%
where $\bar{E}^{\rm inc}$ is the incident electric fields on the crack from the coil and the Ferrite structure, and $\bar{J}^{\rm ind}$ is the induced Eddy current on the crack due to the change of the material within the conductive media \cite{Buvat_AIP_2005}. Eq.~\eqref{Z_variation} facilitates the calculation of the variation of the impedance. The details of computations of the incident fields from the coil with Ferrite core can be found in \cite{Sabbagh_TM_1987}. 

Given the crack profile, the forward problem is to compute the induced Eddy current and then the variation of the impedance according to Eq.~\eqref{Z_variation} \cite{Bowler_TM_1989, Miorelli_TM_2013}. 
We established a forward model based on Eq.~\eqref{Z_variation} which has been verified with data from the literature and from past experiments. It is used to generate simulation data used for training. Without loss of generality, we assume the dimension of all the crack profiles is $40\times 12$ pixels, and the EC responses are mapped to a rectangle of $40\times 40$ pixels.

To generate a crack profile, a binary random matrix of $40\times 12$ is instantiated with equal 0 and 1 probabilities, in which 1's indicate the presence of cracks. However such a profile is highly jagged and fragmented, which is both unrealistic and unfriendly for CNN learning (as CNNs are good at capturing patterns in \emph{smooth} images). Hence the random profiles are smoothed with a median filter of a $3\times 3$ window.

Given a crack profile, the forward model generates simulated EC responses at 3 different frequencies. The responses are complex numbers, whose real part and imaginary part are treated as two separate input channels. Therefore for each profile, the responses can be viewed as a 6-channel image. Fig. \ref{data-sample} shows an example of a random crack profile (after smoothing) and the corresponding EC responses.
\begin{figure*}[th]
\centering
\includegraphics[width=\textwidth]{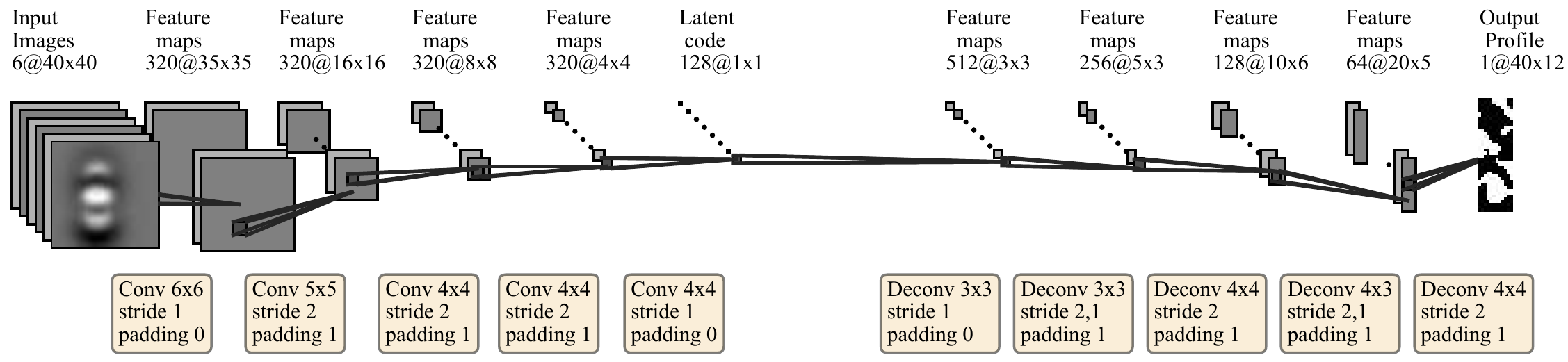}
\caption{The architecture of EddyNet. Each wheat-colored Conv/Deconv block is a convolutional/deconvolutional layer sitting between the left (input) and right (output) feature maps / images.}
\label{arch}
\end{figure*}

\section{EddyNet -- An Encoder-Decoder CNN}
The EddyNet architecture is illustrated in Fig. \ref{arch}. It consists of an encoder (left side) and a decoder (right side). The encoder converts the 6-channel input image to a 128-dimensional latent code, and the decoder decodes the latent code into a single-channel crack profile.

The encoder consists of five convolutional layers. Each of the first four convolutional layer is followed by a BatchNorm layer and a Mish \cite{mish} activation function, not shown in Fig. \ref{arch}. The output feature maps gradually decrease in height and width, until being converted to a 128-dimensional continuous vector, which is referred to as the \emph{latent code} of the input. The first two layers have larger sizes of kernels ($6\times 6$ and $5\times 5$ vs. $4\times 4$ in the last three layers). This design is to equip the lower layers with larger receptive fields, so that they are better at capturing long-range correlations (low spatial frequency signals) across the 2D plane. To preserve maximal information in the first few layers, the numbers of output channels of all layers except for the last are fixed to 320.

The decoder consists of five deconvolutional layers. They gradually upsample the latent code to an output image of size $40\times 12$ as the crack profile. Intuitively this process spreads out the information captured in different channels (corresponding to different patterns in the input image) of the latent code onto the 2D plane. The kernel sizes of these layers tend to increase, so as to fill in long-range and higher-resolution details. Each of the first four deconvolutional layer is followed by a BatchNorm layer and a Mish activation function. The last layer output consists of $K$ channels, which are aggregated with an attention function. The softmax of the channel activations is used as attention weights:
\begin{equation}
    \bar{x}_{ij} = \text{softmax}(\boldsymbol{x}_{\cdot ij})^\intercal \boldsymbol{x}_{\cdot ij},
\end{equation}
where $\boldsymbol{x}_{\cdot ij}$ is the $K$-channel feature vector at pixel $i,j$. Intuitively, the attention mechanism allows more diverse patterns to be stored in the last deconvolutional layer, so that they could be used to reconstruct finer various details of the profile.

Taking 6 channels of EC responses as input, EddyNet outputs an reconstructed crack profile. The mean absolute error (MAE) per pixel between the ground truth and reconstructed profiles is used as the training objective. The motivation of adopting MAE as the metric is that it gives us a quantitative summary of how accurate the reconstruction is over various crack profiles. Admittedly, MAE fails to consider some practically important factors, such as safety implications of different errors. Such metrics should certainly be explored in future studies. Though, we expect the model and training pipeline remain largely the same after incorporating these metrics as loss terms, as long as they are differentiable.

The architecture of EddyNet is largely influenced by the Deep Convolutional Generative Adversarial Networks (DCGAN) \cite{dcgan}, with the discriminator and the generator swapped in the pipeline. However their distinct training objectives differentiate them substantially.

\section{Data Preparation and Model Training}
20,000 pairs of simulated (profile, responses) data were generated. The data was divided into non-overlapping training and test sets with a 80-20\% split.

EddyNet was configured to have $K=20$ channels for the attention function. It was trained with a batch size of 64 for 30 epochs, by the Ranger optimizer \cite{ranger} at an initial learning rate of $0.0002$. The whole training took 8 minutes and 2.4 GBs of GPU RAM on an NVIDIA GeForce Titan X GPU.

\section{Experimental Results}

As an ablation study, we removed the decoder, and instead let the encoder output a 480-dimensional latent code, and reshaped it to a $40\times 12$ crack profile. This model is named \textbf{Eddy-nodec}. The second ablated model, \textbf{Eddy-relu}, was created by replacing the Mish activation functions to LeakyReLU (in the decoder) or to ReLU (in the encoder). The third ablated model, \textbf{Eddy-noattn}, was created by removing the attention mechanism in the decoder.

As pixels in an reconstructed profile are real numbers between $[0,1]$, they were binaried to 0 and 1. On the test set of 4,000 profiles, the MAE of the binarized profiles reconstructed by EddyNet was 0.198. Note that random guesses of the true profiles would yield an MAE of 0.5. This much lower MAE indicates that EddyNet is able to find decent approximate solutions to the inverse problem.
    \begin{figure}[ht]
        \centering 
            \newcolumntype{C}{ >{\centering\arraybackslash} m{3cm} }
\begin{tabular}{CCC}
 \includegraphics[trim=0 169px 0 0,clip,scale=0.8]{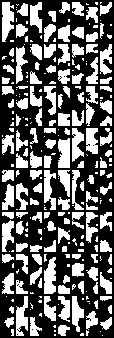} & \includegraphics[trim=0 169px 0 0,clip,scale=0.8]{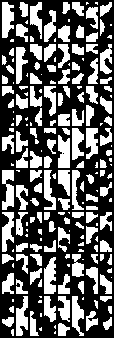} & \includegraphics[trim=0 169px 0 0,clip,scale=0.8]{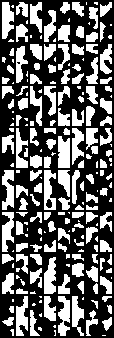} \tabularnewline
 \scriptsize{(a)} & \scriptsize{(b)} & \scriptsize{(c)} 
 \tabularnewline
 \tabularnewline
\includegraphics[trim=0 169px 0 0,clip,scale=0.8]{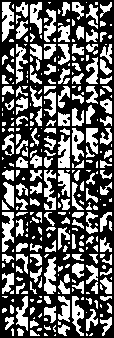} &
\includegraphics[trim=0 169px 0 0,clip,scale=0.8]{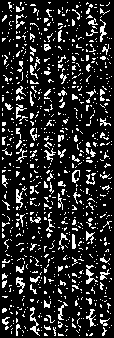} &
\includegraphics[trim=0 169px 0 0,clip,scale=0.8]{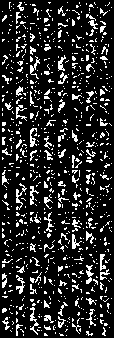}
\tabularnewline
 \scriptsize{(d)} & \scriptsize{(e)} & \scriptsize{(f)}
\tabularnewline
 
\end{tabular}
        \caption{(a) 32 ($4\times 8$) reconstructed profiles by Eddy-nodec; (b) reconstructed profiles by Eddy-relu; (c) reconstructed profiles by EddyNet; (d) ground truth crack profiles; (e) the absolute error between (b) and (d); and (f) the absolute error between (c) and (d).}
        \label{results}
    \end{figure}

For a qualitative analysis, Fig. \ref{results}(d) shows 32 randomly selected true crack profiles from the test set; (a), (b) and (c) are the binarized reconstructed profiles by Eddy-nodec, Eddy-relu and EddyNet, respectively; (e) and (f) show the wrongly reconstructed pixels in (b) and (c), respectively, where a white pixel is wrong and a black pixel is correctly reconstructed.  The more black pixels in (e) or (f), the better the model reconstructs the true profiles. The ground truth and reconstructed crack profiles are visually similar, i.e., containing similar geometries and patterns, although differing in some details. Although achieving a similar MAE, the profiles reconstructed by Eddy-nodec in Fig.~4(a) contained a lot of jagged and fragmented artifacts, which were improved with the extra decoder.

\begin{table}[th]
    \centering
    \caption{MAE achieved by three models.}
    \begin{tabular}{c|c|c|c|c}
    \hline
    profile type & EddyNet & Eddy-nodec & Eddy-relu  & Eddy-noattn     \tabularnewline \hline
        raw        & 0.214 & 0.291  & 0.221 & 0.226 
        \tabularnewline \hline
        binarized  & 0.198 & 0.206 & 0.203 & 0.207 \tabularnewline \hline
    \end{tabular}
    \label{tab:mae}
\end{table}

An advantage of EddyNet for crack profile reconstruction is it runs extremely fast compared to traditional methods. Table \ref{tab:speed} presents the reconstruction time on GPUs and CPUs, respectively.

\begin{table}[th]
    \centering
    \caption{Reconstruction time (sec).}
    \begin{tabular}{c|c|c}
    \hline
    batch size & GPU  & CPU     \tabularnewline \hline
        64     & 0.04 & 0.29    \tabularnewline \hline
        1      & 0.005 & 0.015    \tabularnewline \hline
    \end{tabular}
    \label{tab:speed}
\end{table}

\section{Conclusions}
In this paper, we proposed a neural solver for the inverse problem associated with Eddy current (EC) testing, i.e., reconstructing the crack profile from EC response signals. The inverse mapping from EC responses to crack profiles is learned with \emph{EddyNet}, a deep encoder-decoder convolutional neural network. EddyNet achieved a mean absolute error of 0.198 on the held-out test data. The reconstructed crack profiles are visually similar to the true profiles. In addition, EddyNet only takes $5\sim 15$ milliseconds to reconstruct a crack profile. As EddyNet is a generic framework, it has potential applications on various inverse problems beyond crack profile reconstruction.

Our future work will focus on investigating EddyNet in more practical settings. One thing worth exploring is improve random crack profile generation methods, to make the random profiles physically more realistic. Another future topic is use domain-specific metrics of reconstruction errors to guide the model training, such as those considering safety implications of reconstruction errors.


%

\end{document}